\documentclass[twocolumn,printnumbers,amsmath,amssymb,prl]{revtex4}
\usepackage{graphicx}% Include figure files
\usepackage{color}

\begin{document}
\title{Realizing negative Poisson's ratio in unstressed spring networks}

\date{\today}

\author{Jun Liu$^{1}$}
\author{Yunhuan Nie$^1$}
\author{Hua Tong$^2$}
\author{Ning Xu$^{1,*}$}

\affiliation{$^1$Hefei National Laboratory for Physical Sciences at the Microscale, CAS Key Laboratory of Soft Matter Chemistry, and Department of Physics, University of Science and Technology of China, Hefei 230026, P. R. China \\$^2$Department of Fundamental Engineering, Institute of Industrial Science, University of Tokyo, 4-6-1 Komaba, Meguro-ku, Tokyo 153-8505, Japan}

\begin{abstract}
When randomly displacing the nodes of a crystalline and unstressed spring network, we find that the Possion's ratio decreases with the increase of structural disorder and even becomes negative. Employing our finding that longer springs tend to contribute more to the shear modulus but less to the bulk modulus, we are able to achieve negative Poisson's ratio with lower structural disorder by attributing each spring a length dependent stiffness. Even with perfect crystalline structure, the network can have negative Possion's ratio, if the stiffness of each spring is set by its virtual length after a virtual network distortion. We also reveal that the nonaffine contribution arising from the structural or spring constant disorder produced in some cooperative way by network distortion is essential to the emergence of negative Poisson's ratio.
\end{abstract}

\maketitle

When an elastic material is stretched or compressed in one direction by a strain of $\epsilon_{\parallel}$, it deforms by a strain of $\epsilon_{\perp}$ in the perpendicular direction. Such a response is characterized by the Poisson's ratio \cite{landau},
\begin{equation}
\nu = - {\rm d}\epsilon_{\perp} / {\rm d}\epsilon_{\parallel}. \label{eq1}
\end{equation}
Everyday experience leads us to believe that normal materials should have positive Poisson's ratio, as a rubber band does. However, people have achieved negative Poisson's ratio by designing or synthesizing materials with specific microstructures \cite{lakes,gardner,evans,greaves,baughman,larsen,jiang,schenk,wei,grima}. These artificial materials, named as auxetic materials \cite{evans}, form a family of mechanical metamaterials with important applications \cite{greaves}.

For isotropic materials, the Poisson's ratio can be also expressed in terms of elastic moduli \cite{greaves}:
\begin{equation}
\nu_{iso} = \frac{d-2G/B}{d(d-1)+2G/B}, \label{eq2}
\end{equation}
where $d$, $G$ and $B$ are the dimension of space, shear modulus and bulk modulus, respectively. In order to realize negative Poisson's ratio, the material is required to have a large enough $G/B(> d/2)$.  This provides a way to design auxetic materials by tuning material's elastic moduli, although it is challenging and even counter-intuitive to have isotropic materials to be rigid to shear but vulnerable to compression. Recent studies have suggested that disordered solids are good isotropic systems to start with \cite{goodrich,reid,hexner,rocks,yan}. At the bond level, responses to compression and shear in disordered solids are independent, so that negative Poisson's ratio is realizable by selectively pruning bonds \cite{goodrich}.

In this letter, we report simple and practical realizations of auxetic spring networks, in usage of disorder as well, but not limited to structural disorder.  By distorting an unstressed spring network from the initial crystalline structure, we observe that the Poisson's ratio decreases and surprisingly drops below zero with the increase of structural disorder characterized by the magnitude of node displacement $\eta$.  We find that longer springs tend to contribute more to the shear modulus but less to the bulk modulus. By distributing each spring a length dependent stiffness, we realize negative Poisson's ratio at smaller $\eta$, even in the $\eta=0$ limit where the network maintains perfect crystalline structure.  Our further analyses reveal that negative Poisson's ratio is induced by strong nonaffinity. The way that we distort the network and manipulate the spring constant efficiently enhances the nonaffinity due to some elusive underlying mechanisms.

Here we concentrate on two-dimensional systems with periodic boundary conditions in both directions. We have verified that our major findings hold for three-dimensional systems as well. We start with a perfect crystalline lattice with nearest lattice sites (nodes) being connected by springs of length $l_0$. Structural disorder is introduced by randomly displacing node $i$ ($i=1,2,...,N$) from crystalline lattice site $\vec{r}_{i,c}$ to $\vec{r}_i=\vec{r}_{i,c}+\vec{\eta}_i$, where $\vec{\eta}_i=(\eta{\rm cos}\theta_i, \eta{\rm sin}\theta_i)$ with $\eta\in [0, l_0/2]$ being constant for all nodes and $\theta_i$ being a random angle ranging from $0$ to $2\pi$.  The topology of the network, i.e., number of springs and their connections, remain unchanged after the distortion. To maintain mechanical stability, all networks concerned here are unstressed with all springs being relaxed. Normally, larger $\eta$ causes stronger structural disorder.  We set $\eta\le l_0/2$ in order to avoid crossing of springs. In practice, we generate a random set of angles $\theta_i$ initially and then grow $\eta$ while fixing the angles. Data are averaged over different realizations of the angles.

The elastic moduli and Possion's ratio are obtained by slightly deforming the unstressed networks and calculating the linear response after the energy minimization \cite{fire}. Upon deformation, the energy stored in the spring connecting nodes $i$ and $j$ is
\begin{equation}
U_{ij} = \frac{1}{2}k_{ij}\left( r_{ij}- l_{ij}\right)^2, \label{eq:pot_net}
\end{equation}
where $k_{ij}$ is the spring constant, $r_{ij}$ is the separation between nodes (i.e., length of the spring upon deformation), and $l_{ij}$ is the length of relaxed spring before deformation. We set the crystalline lattice constant $l_0$ to be one. In the first part of this letter, we discuss effects of structural disorder and set all $k_{ij}$ to be unity.  Afterwards, we  will vary the spring constants following the rules to be introduced.

%%%%%%%%%%%%%%%%%%%%%%%%%%%%%%%%%%%%%%%%%%%%%%%%%%%
\begin{figure}
\vspace{-0.13in}
\includegraphics[width=0.48\textwidth]{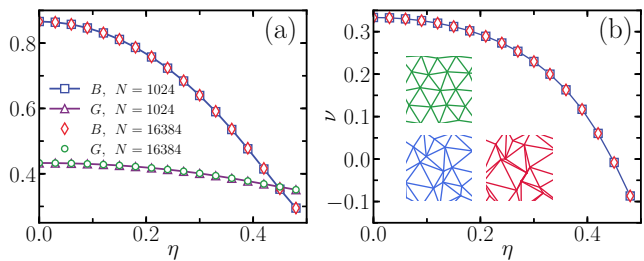}
\caption{\label{fig:fig1} Elastic properties subject to the network distortion from a triangular lattice with all springs having the same stiffness. (a) Shear and bulk moduli, $G$ and $B$, and (b) Possion's ratio $\nu$ against the magnitude of node displacement $\eta$. $\eta=0$ corresponds to a perfect triangular lattice. Squares and diamonds in (b) are $\nu(\eta)$ from Eq.~(\ref{eq1}) and $\nu_{iso}(\eta)$ from Eq.~(\ref{eq2}), respectively, for $N=1024$ systems. The lines are guides for the eye. The insets of (b) are parts of the networks at $\eta=0.1$ (top), $0.3$ (left bottom), and $0.46$ (right bottom).
	}
\end{figure}
%%%%%%%%%%%%%%%%%%%%%%%%%%%%%%%%%%%%%%%%%%%%%%%%%%%

Figure~\ref{fig:fig1} shows results of spring networks distorted from a triangular lattice. Each node is connected by $6$ springs, so the average coordination number $z$ is equal to $6$, well above the isostatic value $z_{iso}=2d=4$. This guarantees the mechanical stability of the unstressed spring networks \cite{ohern,silbert,wyart,mao,nie,tong,wang}.

Figure~\ref{fig:fig1}(a) compares the bulk and shear moduli \cite{note2}. The data collapse of $N=1024$ and $16384$ systems confirms that our results do not rely on system size. Both $G$ and $B$ decrease when $\eta$ increases. Apparently, $B$ decays more quickly than $G$. Surprisingly, $G=B$ at a critical value of the distortion, $\eta_c\approx 0.446$. When $\eta >\eta_c$, $B$ is even smaller than $G$. From Eq.~(\ref{eq2}), negative Poisson's ratio is expected at $\eta>\eta_c$.

We then calculate the Poisson's ratio $\nu$ according to Eq.~(\ref{eq1}). Because our networks are unstressed, we apply a small compressive strain $\epsilon_{\parallel}$ in the $x$-direction and minimize the energy while maintaining zero pressure in the $y$-direction. The network responds to the compression by a deformation strain $\epsilon_{\perp}$ in the $y$-direction. As long as $\epsilon_{\parallel}$ is small, we can always obtain a linear response, from which we determine $\nu$ from Eq.~(\ref{eq1}). As shown in Fig.~\ref{fig:fig1}(b),  $\nu$ decreases when $\eta$ increases, which becomes negative when $\eta>\eta_c$. We also show $\nu_{iso}$ calculated from Eq.~(\ref{eq2}), which collapses well with $\nu$.

Since negative Poisson's ratio emerges when $\eta$ is large, disorder should be essential to the formation of auxetic networks. There are different ways to introduce disorder to a triangular lattice, e.g., randomly distorting nodes as done here, removing bonds, removing nodes, or varying bond stiffness \cite{nie}. However, we only obtain negative $\nu$ by distorting nodes. This implies that disorder needs to be introduced in some cooperative other than totally random way, while distorting nodes happens to create such cooperation. The direct consequence of distorting nodes is the cause of spring length disorder. It is thus straightforward to speculate whether there are any underlying connections between spring length and elastic moduli.

%%%%%%%%%%%%%%%%%%%%%%%%%%%%%%%%%%%%%%%%%%%%%%%%%%%
\begin{figure}
\includegraphics[width=0.48\textwidth]{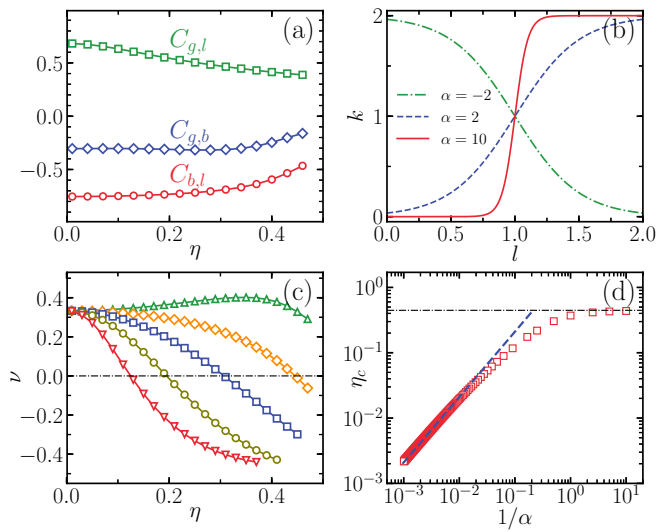}
\caption{\label{fig:fig2} Effects of spring constant disorder on elastic properties of the networks. (a) Correlation functions $C_{g,l}$, $C_{g,b}$, and $C_{b,l}$ defined in the text against the magnitude of node displacement $\eta$ for $N=4096$ systems. (b) Examples of spring constant function $k(l)=1+{\rm tanh}[\alpha(l-l_0)]$ with $l_0=1$ being the unit of length. (c) Poisson's ratio $\nu$ against $\eta$ when spring constants are set by $k(l)$. From left to right, $\alpha=10$ (down triangles), $5$ (circles), $2$ (squares), $0$ (diamonds), and $-2$ (up triangles). The solid lines are guides for the eye. The horizontal dot-dashed line labels $\nu =0$. (d) Correlation between $\eta_c$, the critical value of $\eta$ at which $\nu=0$, and $\alpha^{-1}$, the ``'theoretical' width of spring constant according to $k(l)$. The dashed line shows the linear relation $\eta_c\sim \alpha^{-1}$ in the large $\alpha$ limit. The horizontal dot-dashed line labels the upper bound $\eta_c\approx 0.446$ when $\alpha=0$.
	}
\end{figure}
%%%%%%%%%%%%%%%%%%%%%%%%%%%%%%%%%%%%%%%%%%%%%%%%%%%

Inspired by the approach of decomposing elastic moduli into bonds: $G=\sum_i g_i$ and $B=\sum_i b_i$ with the sums being over all bonds \cite{goodrich}, we calculate $g_i$ and $b_i$,  the contributions of spring $i$ to the shear and bulk moduli, for all $3N$ springs. Figure~\ref{fig:fig2}(a) shows that $C_{g,b}=\left<\delta g\delta b\right>=\left< (g -\bar{g})(b-\bar{b})\right><0$ when $\eta>0$, where $\bar{g}=G/3N$, $\bar{b}=B/3N$, and $\left<.\right>$ denotes the average over springs and configurations. Therefore, $\delta g$ and $\delta b$ are negatively correlated, implying that a spring which helps to strengthen $G$ tends to weaken $B$. Figure~\ref{fig:fig2}(a) also shows that $C_{g,l}=\left<\delta g\delta l\right>>0$ and $C_{b,l}=\left<\delta b \delta l\right><0$, where $\delta l = l - \bar{l}$ with $l$ and $\bar{l}$ being the spring length and its average value. These correlations hint that longer springs ($l>\bar{l}$) tend to have larger $g$ and smaller $b$. This finding is rather inspiring. Up to now, all springs are set to have identical spring constant. If we make longer springs stiffer and shorter ones softer, we may be able to enhance $G$ and meanwhile weaken $B$, so that the Poisson's ratio may become negative at smaller $\eta$.

We  then let the spring constant $k$ [$k_{ij}$ in Eq.~(\ref{eq:pot_net})] be an increasing function of $l$. By employing various functional forms of $k(l)$, we find that such manipulation of $k$ indeed significantly decreases $\nu$. Here we present results for a chosen form $k(l)=1+{\rm tanh}[\alpha (l - l_0)]$, where $\alpha$ sets the steepness of $k(l)$ in the vicinity of the triangular lattice constant $l_0$. As shown in Fig.~\ref{fig:fig2}(b), $k(l)$ is bounded in $[0,2]$. The results in Fig.~\ref{fig:fig1} are just for the special case of $\alpha=0$.

For a given $\alpha$, we distort the perfect triangular lattice as done above for $\alpha=0$ and simultaneously distribute each spring a spring constant $k(l)$. For all $\alpha >0$, Fig.~\ref{fig:fig2}(c) shows that $\nu(\eta)$ behaves similarly to the $\alpha=0$ case. Interestingly, the critical value of the distortion $\eta_c$ at which $\nu=0$ decreases with the increase of $\alpha$.  The involvement of spring constant disorder associated with the spring length greatly weakens the required structural disorder for auxetic networks to occur. For comparison, we also show in Fig.~\ref{fig:fig2}(c) an example with $\alpha<0$. As expected, negative $\alpha$ hinders the decrease of $\nu$ and pushes $\eta_c$ to higher values. With sufficiently large $|\alpha|$, $\nu$ even grows with $\eta$. By forcing longer springs to be softer, we instead are able to achieve much larger $\nu$ than that for perfect crystalline lattice.  The manipulation of $\alpha$ thus offers us the freedom to control the elastic moduli and Poisson's ratio purposely.

Figure~\ref{fig:fig2}(d) shows how the critical structural disorder $\eta_c$ varies with $\alpha$. As mentioned above, when $\alpha=0$, $\eta_c\approx 0.446$, which sets the upper bound of $\eta_c$ purely induced by structural disorder. When $\alpha$ is large, Fig.~\ref{fig:fig2}(d) indicates that $\eta_c\sim \alpha^{-1}$. Negative Poisson's ratio there is mostly contributed by spring constant disorder, because the structural disorder is too small to cause significant change of $\nu$ on its own, as shown by curves in Fig.~\ref{fig:fig1} in the $\eta\rightarrow 0$ limit. The structural disorder just acts to provide a way to cause some cooperative spring constant distribution. Note that $\alpha^{-1}$ sets the ``theoretical'' width of spring constant distribution. The linear relation between $\eta_c$ and $\alpha^{-1}$ indicates that, when spring constant disorder dominates, the structural disorder must lead to a spring constant distribution comparable to $\alpha^{-1}$ in order for auxetic networks to occur. When $\alpha$ is small, both structural and spring constant disorder contribute to the formation of auxetic networks, so $\eta_c(\alpha^{-1})$ deviates from the linear relation and approaches the upper limit gradually.

Figure~\ref{fig:fig2}(d) implies that negative Possion's ratio is possible in perfect crystalline lattices, if we were able to perform an $\eta\rightarrow 0$ distortion and set the spring constants to be either $2$ or $0$, because $\alpha\rightarrow \infty$. This implies that auxetic networks can also be produced by removing bonds of a perfect lattice following the $\eta\rightarrow 0$ distortion, as long as the resulting network is still mechanically stable. Next, however, we will show that there is an alternative scheme to realize negative Poisson's ratio in the $\eta=0$ networks without destroying the network topology.

%%%%%%%%%%%%%%%%%%%%%%%%%%%%%%%%%%%%%%%%%%%%%%%%%%%
\begin{figure}
\includegraphics[width=0.48\textwidth]{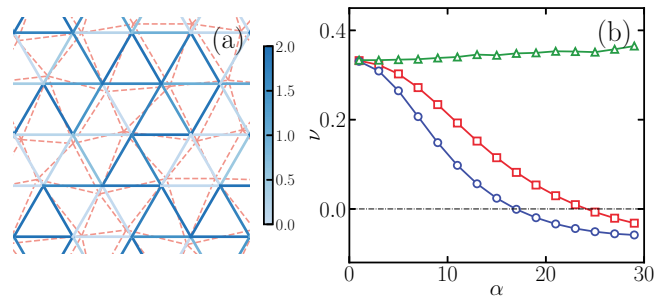}
\caption{\label{fig:fig3} Elastic properties of the perfect triangular lattice with spring constant disordered introduced by virtual network distortion. (a) Illustration of a part of a triangular lattice (solid) and its virtual distortion (dashed). The gray scale of the solid lines shows the value of spring constant determined by the virtual network distortion ($\eta^*=0.15$ and $\alpha=10$), as quantified by the scale bar. (b) Possion's ratio $\nu$ against $\alpha$ for the $N=16384$ triangular lattice with spring constants being set by a virtual network distortion of $\eta^*=0.1$ (squares) and $0.15$ (circles). The triangles show the case when spring constants set by $\eta^*=0.15$ are randomly distributed to bonds. The solid lines are guides for the eye. The horizontal dot-dashed line labels $\nu=0$.
	}
\end{figure}
%%%%%%%%%%%%%%%%%%%%%%%%%%%%%%%%%%%%%%%%%%%%%%%%%%%

An interesting and inspiring observation from Fig.~\ref{fig:fig2}(d) is that $\alpha\sim \eta_c^{-1}$ when $\alpha$ is large. Note that here $\eta_c$ corresponds to a real network distortion, but such a small structural disorder does not actually affect $\nu$ and is negligible. Then, if we instead assume a virtual distortion $\eta^*$ to a perfect lattice and distribute each bond a spring constant $k(l^*)$ with $l^*$ being the virtual length of the bond after the virtual distortion, as illustrated by Fig.~\ref{fig:fig3}(a), can we achieve auxetic $\eta=0$ networks with finite $\alpha$?

Figure~\ref{fig:fig3}(b) shows that the virtual distortion scheme indeed works. At fixed $\eta^*$, $\nu$ decreases with the increase of $\alpha$ and even drops below zero. With larger $\eta^*$, smaller $\alpha$ is required to obtain negative Poisson's ratio. Different from most of previous approaches, auxetic materials are obtained here without designing any specific structures or destroying network topology, but just by utilizing spring constant disorder.  To highlight the importance of cooperative spring constant distribution generated by the lattice distortion, we shown in Fig.~\ref{fig:fig3}(b) another $\nu(\alpha)$ curve obtained by randomly distributing the same set of spring constants from virtual distortion. In sharp contrast, the Poisson's ratio does not decay with the increase of $\alpha$ any more. 

We have seen that disorder is necessary for auxetic networks to occur, although here it is not introduced in a totally random way. The presence of disorder and the interplay between two types of disorder make it difficult to pin down whether there is any geometric origin of auxetic networks.  However, to study some unique properties induced by disorder may shed some light on our understanding of the formation of auxetic networks.

One direct consequence of disorder is the nonaffine deformation subject to strain \cite{zaccone,ellenbroek,maloney}. In order to quantify the effects of nonaffinity, we calculate $R={\rm d}P_{\perp}/{\rm d}\epsilon_{\parallel}$ and decompose it into an affine part $R_a$ and a nonaffine part $R_{na}$ \cite{zaccone,ellenbroek,maloney}:
\begin{equation}
R = R_a + R_{na} = -\frac{1}{A} \frac{\partial^2 U}{\partial\epsilon_{\parallel}\partial \epsilon_{\perp}} - \frac{1}{A}\sum_i\frac{\partial^2 U}{\partial \epsilon_{\parallel}\partial \vec{r}_i}\cdot \frac{{\rm d}\vec{r}_i}{{\rm d}\epsilon_{\perp}},
\end{equation}
where $P_{\perp}$ is the pressure in the $y$-direction induced by the strain $\epsilon_{\parallel}$ in the $x$-direction, $U$ is the total potential energy, $A$ is the area of the system, and the sum is over all nodes. $R_{na}$ can be calculated from the inverse of Hessian matrix \cite{zaccone,ellenbroek,maloney}. Apparently, $R$ is negative when $\nu<0$.

%%%%%%%%%%%%%%%%%%%%%%%%%%%%%%%%%%%%%%%%%%%%%%%%%%%
\begin{figure}
\includegraphics[width=0.48\textwidth]{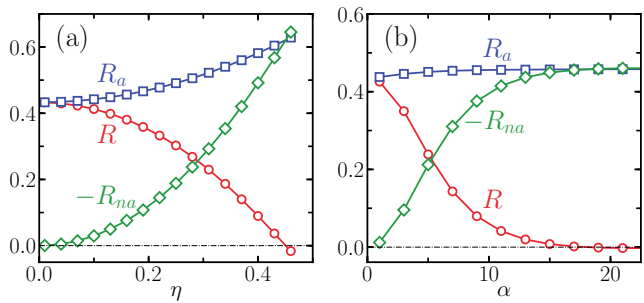}
\caption{\label{fig:fig4} Comparison of the total response $R$ and its affine and nonaffine components $R_a$ and $R_{na}$ as defined in the text. $R_{na}$ is always negative. In order to quantitatively compare it with $R_a$, which is positive, here we show $-R_{na}$. (a) and (b) are for the cases shown in Figs.~\ref{fig:fig1}(b) and \ref{fig:fig3}(b) ($\eta^*=0.15$), respectively. The horizontal dot-dashed lines label $R=0$.
	}
\end{figure}
%%%%%%%%%%%%%%%%%%%%%%%%%%%%%%%%%%%%%%%%%%%%%%%%%%%

In Fig.~\ref{fig:fig4}, we compare $R$, $R_a$, and $R_{na}$ for the $\alpha=0$ and $\eta=0$ ($\eta^*=0.15$) cases. The affine component $R_a$ is always positive and does not decay with the increase of $\eta$ or $\alpha$, which is apparently harmful to the formation of negative $\nu$. In contrast, $R_{na}$ is negative, whose absolute value increases when $\eta$ or $\alpha$ increases and eventually beats $R_{a}$. Therefore, nonaffinity induced by disorder is responsible to the formation of auxetic networks. However, nonaffinity is ubiquitous in disordered solids, while most of disordered solids do not have negative Poisson's ratio. The way that we produce auxetic networks happens to generate some elusive mechanisms to boost the contribution of nonaffinity, which should be different from normal disordered solids and calls for future studies.

Taking advantage of disorder, we propose experimentally accessible methods to tune elastic moduli and obtain auxetic materials. The key is that the disorder must be introduced in some cooperative way induced by (virtual) network distortion. Here we only show results of manipulating triangular lattice. Our major findings should be general and valid to other types of lattices. However, more work is required to figure out whether and how lattice topology may affect the results, especially for some special ones like kagome lattice \cite{mekata} which already exhibit extraordinary elastic properties \cite{sun,mao1}.

An important finding of our work is that the Poisson's ratio can be tuned by adjusting the bond stiffness, without any change of the structure. Most of previous studies on auxetic materials have been focused on structures. Here we show that structure is not the only aspect that we should consider in designing auxetic materials, because we have achieved auxetic materials even with pure crystalline structures.  However, as shown in Fig.~\ref{fig:fig3}(b), randomly adjusting the bond stiffness does not effectively affect the Poisson's ratio. Our success is based on the observation of the correlations between elastic moduli and bond length associated with the network distortion.

By manipulating bond stiffness, we obtain the freedom to tune elastic properties of possibly any unstressed stable networks. In addition to auxetic materials, the $\alpha<0$ curve in Fig.~\ref{fig:fig2}(c) indicates that the Poisson's ratio of networks with apparent structural disorder can be tuned to that of crystals. We expect that adjusting bond stiffness of unstressed disordered solids using $\alpha<0$ may induce extraordinary mechanical and vibrational properties distinct from normal disordered solids. This may reveal new aspects of disordered solids and help us further understand how different types of disorder compete or assist each other to determine special properties of disordered solids.

We thank Peng Tan and Fangfu Ye for instructive discussions.  This work was supported by the National Natural Science Foundation of China Grants No.~11734014 and No. 11574278. We also thank the Supercomputing Center of University of Science and Technology of China for the computer time.


\begin{thebibliography}{50}
\bibitem{note2} For triangular lattice which is actually anisotropic, the shear modulus calculated from the simple shear is equal to that from the pure shear, so we do not need to worry about the variation of the shear modulus due to the ways to impose shear.
\end{thebibliography}
\end{document}